\documentclass[a4paper,twocolumn,english,nofootinbib,pre,superscriptaddress,aps]{revtex4-2}
\usepackage{amsmath,amsfonts,amsthm,bm,physics}
\usepackage{amssymb}
\usepackage{mathrsfs}
\usepackage{graphicx}
\usepackage{xcolor}
\usepackage{hyperref}
\usepackage{xr-hyper}

\makeatletter
\newcommand*{\addFileDependency}[1]{
\typeout{(#1)}
\@addtofilelist{#1}
\IfFileExists{#1}{}{\typeout{No file #1.}}
}
\makeatother

\newcommand*{\myexternaldocument}[1]{
\externaldocument{#1}
\addFileDependency{#1.tex}
\addFileDependency{#1.aux}
}

\myexternaldocument{SupM}

\newcommand{\ra}{\rangle}

\newcommand{\ww}{\widetilde}

\begin{document}

\title{
 An operator-Weyl-symbol approach to eigenstate thermalization hypothesis}

\author{Xiao Wang}\email{wx2398@ustc.edu.cn}
\affiliation{ Department of Modern Physics, University of Science and Technology of China,
	Hefei 230026, China}
\affiliation{CAS Key Laboratory of Microscale Magnetic Resonance,
University of Science and Technology of China, Hefei 230026, China}
\author{Wen-ge Wang}\email{wgwang@ustc.edu.cn}
\affiliation{ Department of Modern Physics, University of Science and Technology of China,
	Hefei 230026, China}
\affiliation{CAS Key Laboratory of Microscale Magnetic Resonance,
University of Science and Technology of China, Hefei 230026, China}
\affiliation{Anhui Center for fundamental sciences in theoretical physics, Hefei 230026, China}

\date{\today}

\begin{abstract}
  In this letter, by an approach that employs Weyl symbols for operators,
  a semiclassical theory is developed
  for the offdiagonal function in the eigenstate thermalization hypothesis,
 which is for offdiagonal elements $\langle{E_i}\left|O\right|{E_j}\rangle$
 of an observable $O$ on the energy basis.
 It is shown analytically that the matrix of $O$ has a banded structure,
 possessing a bandwidth $w_b$ that scales linearly with $\hbar$, a phase-space gradient of the
  classical Hamiltonian, $\langle\left|{\boldsymbol{\nabla }H_{\rm cl}}\right|\rangle$, and an
  $O$-dependent property.
 This predicts that the thermalization timescale of a quantum system may be
 inversely proportional to the phase-space gradient of the Hamiltonian, aligning with
 intuitions in classical thermalization.
 This approach also elucidates the origin of a $\rho_{\rm dos}^{-1/2}$-scaling of the offdiagonal function.
 The analytical predictions are checked numerically in the Lipkin-Meshkov-Glick model.
\end{abstract}

\maketitle


The eigenstate thermalization hypothesis (ETH) provides a fundamental way of
understanding how isolated quantum many-body systems approach thermal equilibrium
\cite{Deutch91,PRA86-Feing-Peres,srednicki1994chaos,srednicki-JPA96,srednicki1999approach,Rigol-AiP16,Gogolin_2018,Turner_2018,Vidmar_2019,Vidmar21-PRB,Rigol_2021,Rigol_2020,Rigol_2017,Rigol_2019,Gemmer_2020,Jiaozi_2025,Vidmar_2024,Jafferis_2023,PSSV-RMP11,EisertFG15-NP,Tasaki16-typi-therm,GE16-thermal-review,BISZ-PR16,Mori-IKU17-review}.
It posits that the matrix elements of a physical observable $O$ in the energy eigenbasis
$\{\ket{E_i}\}$ take a specific form:
\begin{equation}
  \bra{E_i}O\ket{E_j}=O(e)\delta_{ij}+\rho_{\rm dos}^{-1/2}\left(e\right)f(e,\omega)r_{ij},
\end{equation}
where $e=(E_i+E_j)/2$, $\omega=E_i-E_j$, $\rho_{\rm dos}$ is the density of states, $O(e)$
and $f(e,\omega)$ are smooth functions, and $r_{ij}=r^*_{ij}$ are assumed to be random variables with normal distribution (zero
mean and unit variance).

During the past three decades since ETH was proposed, numerous outstanding experimental
and numerical studies have been conducted \cite{Rigol-AiP16,Deutch-RPP18,RS-PRL12,PSGC-PRL15,Deutch-RPP18,Bernien_2017,
Turner_2018,Rigol_2008,Garrison_2018,Srednicki_Rigol_2013,Rigol_2021,Rigol_2020,Steinigeweg_2013,
Rigol_2017,Vidmar_2019,Rigol_2019,Gemmer_2020,WangXiao_2024}. The validity of ETH has
been numerically confirmed for a variety of models and observables \cite{Rigol_2010,Beugeling_2015,Beugeling_2014,Vidmar_2019,Rigol_2019,Rigol_2017,YWW2022PS,Steinigeweg_2013}. Generally, the ETH
is believed to hold in (at least) nonintegrable models.
However, despite significant progresses achieved in ETH research,
 a deep analytical understanding of the offdiagonal
function $f(e,\omega)$ — crucial for characterizing eigenstate correlations \cite{Dominik_2024,Pappalardi_2022,Pappalardi_2025,WangJiaozi_2022,WangXiao_2024,WangXiao_CTP_2025},
thermalization timescales \cite{Jiaozi_2022,Schiulaz_2019,Serbyn_2017,Prosen_Vidmar_2020,Prosen_2021},
etc — has remained elusive.

In this letter, we bridge this gap by combing a generalization of Wigner function to operators
and a semiclassical approach to the offdiagonal function. Although semiclassical methods
have long been used in the discussion of the diagonal function $O(e)$ \cite{srednicki-JPA96,PRA86-Feing-Peres,WangXiao_2024},
a successful semiclassical approach to the offdiagonal function is still lacking. Our
approach is based on an exact expression for the offdiagonal function,
which makes use of the Weyl symbol of operator,
together with the so-called Berry's conjecture.
It naturally predicts a banded structure for the offdiagonal function, consistent with
numerical observations \cite{YWW2022PS,WangDingZu_2024,Vidmar_2019,Rigol_2017}.
And it yields a first analytical expression for the bandwidth $w_b$ of this band structure.
Besides, the semiclassical expression of offdiagonal function also naturally explains
the characteristic $\rho_{\rm dos}^{-1/2}$ scaling of $\bra{E_i}O\ket{E_j}$. The
analytical predictions are verified numerically in the chaotic regime of the
Lipkin-Meshkov-Glick (LMG) model.

Moreover, this framework turns out fruitful in providing deeper understanding for
thermalization properties of many-body quantum chaotic systems. For example, it reveals that the
thermalization timescale should be inversely proportional to the phase-space
gradient of the Hamiltonian,
in consistency with the well know fact that low particle velocity and weak
interaction, which represent slow particle diffusion and slow energy transport,
respectively, should result in slow thermalization.
In addition, it also reveals the thermalization time of an
open quantum system should depend neither on size of the (huge) environment, 
nor on details of the internal interactions within the environment.

\paragraph*{{Notation.}}
We begin with introducing key concepts and notations essential to our discussions.
The quantum momentum and position operators are denoted by
$\hat{\bm{p}}$ and $\hat{\bm{q}}$ with eigenstates $\ket{\bm{p}}$ and $\ket{\bm{q}}$, respectively,
meanwhile, their classical counterparts are written as $\bm{p}$ and $\bm{q}$.
 The Hamiltonian of the system is written as $H(\hat{\bm{p}},\hat{\bm{q}})$,
 with eigenstates $|E_i\ra$ in the energy order, and an observable to be studied
 is written as $O(\hat{\bm{p}},\hat{\bm{q}})$.
Their classical counterparts, indicated as $H_{\rm cl}(\bm{p},\bm{q})$ and
$O_{\rm cl}(\bm{p},\bm{q})$, are gotten by substituting the classical variables
$(\bm{p},\bm{q})$ into the quantum expressions in place of $(\hat{\bm{p}},\hat{\bm{q}})$.
 The Wigner function for a state $\ket{E_i}$ is written as
\begin{equation}\label{WignerFunction_Definition}
  W_i(\bm{p},\bm{q})=\frac{1}{(2\pi\hbar)^d}\int d\bm{r}
  \ \psi_i^*(\bm{q}+\frac{\bm{r}}{2})\psi_i(\bm{q}-\frac{\bm{r}}{2})
  e^{i\bm{p}\cdot\bm{r}/\hbar},
\end{equation}
where $d$ is the dimension of the configuration space and $\psi_i(\bm{q})=\braket{\bm{q}}{E_i}$
represents the energy eigenfunctions in the configuration space.

 Our discussions are to be based on the following exact expression for the offdiagonal matrix elements 
 ($i\neq j$)
\footnote{A detailed derivation of Eq.(\ref{Oij_WignerJ}) is provided in Sec.\ref{Derivation_Oij_WignerJ} of the
Supplementary Materials.},
\begin{equation}\label{Oij_WignerJ}
  \begin{aligned}
    &\abs{\bra{E_i}O\ket{E_j}}^2
    =\int d\bm{p} d\bm{q} \int d\bm{p}' d\bm{q}'\
    J\left(\bm{p},\bm{q},\bm{p}',\bm{q}'\right)\\
    &\quad \times W_i\left(\bm{p}-\frac{\bm{p}'}{2},\bm{q}-\frac{\bm{q}'}{2}\right)
    W_j\left(\bm{p}+\frac{\bm{p}'}{2},\bm{q}+\frac{\bm{q}'}{2}\right),
  \end{aligned}
\end{equation}
where $J$ is a \textit{bilinear phase-space convolution} of Weyl symbols, which is defined by
\begin{equation}\label{J_Definition}
  \begin{aligned}
    J&\left(\bm{p},\bm{q},\bm{p}',\bm{q}'\right)
    =\frac{1}{(2\pi\hbar)^d}\int d\widetilde{\bm{p}} d\widetilde{\bm{q}}
    \ e^{-i\widetilde{\bm{p}}\cdot\bm{q}'/\hbar}e^{-i\bm{p}'\cdot\widetilde{\bm{q}}/\hbar}\\
    &\times O_{\rm w}\left(\bm{p}+\frac{1}{2}\widetilde{\bm{p}},\bm{q}-\frac{1}{2}\widetilde{\bm{q}}\right)
    O_{\rm w}\left(\bm{p}-\frac{1}{2}\widetilde{\bm{p}},\bm{q}+\frac{1}{2}\widetilde{\bm{q}}\right).
  \end{aligned}
\end{equation}
Here, $O_{\rm w}(\bm{p},\bm{q})$ is the \textit{Weyl symbol} \cite{Weyl_1927} of the operator $O(\hat{\bm{p}},\hat{\bm{q}})$,  defined by
\begin{equation}\label{Osc_Definition}
  O_{\rm w}(\bm{p},\bm{q})
  =\int d\bm{r}\ e^{-i \bm{p}\cdot\bm{r}/\hbar}
  \bra{\bm{q}+\frac{\bm{r}}{2}}O(\hat{\bm{p}},\hat{\bm{q}})\ket{\bm{q}-\frac{\bm{r}}{2}}.
\end{equation}
 The explicit expression of $O_{\rm w}(\bm{p},\bm{q})$ is obtainable by
 writing the operator function $O(\hat{\bm{p}},\hat{\bm{q}})$ as a sum of Weyl-ordered functions\footnote{
 See Sec.\ref{Osc_Introduction} of the Supplementary Materials for
 discussions on properties of $O_{\rm w}$.}
 \cite{WangXiao_2024},
 giving that
\begin{equation}\label{Osc-Ocl}
  O_{\rm w}(\bm{p},\bm{q}) = O_{\rm cl}(\bm{p},\bm{q}) + {\rm O}(\hbar^m) \quad (m \ge 1).
\end{equation}
 Below, we discuss the case of small $\hbar$, in which $O_{\rm w} \approx O_{\rm cl}$.

\paragraph*{{Band Shape.}}
 Equation (\ref{Oij_WignerJ}) predicts a banded shape of the  matrix of ${\bra{E_i}O\ket{E_j}}$.
 To show this point,
 one may make use of Berry's conjecture on the averaged Wigner function,
 which is semiclassically  approximated by
 the energy surface in phase space under $H_{\rm cl}(\bm{p},\bm{q})=E_i$
\cite{Berry77,Berry_1991,Voros_1976,Voros_1977,pre18-EF-BC},
\begin{equation}\label{Wigner_Semiclassical}
 \overline{W}_i(\bm{p},\bm{q})\approx \frac{\delta(H_{\rm cl}(\bm{p},\bm{q})-E_i)}{S(E_i)},
\end{equation}
where $S(E_i)$ indicates the area of the energy surface,
\begin{equation}\label{S_AreaofIsosurface}
  S(E_i)=\int d\bm{p}d\bm{q}\ \delta(H_{\rm cl}(\bm{p},\bm{q})-E_i),
\end{equation}
and the average is taken over a narrow energy shell. Under Eq.(\ref{Wigner_Semiclassical}),
the integrand in Eq.(\ref{Oij_WignerJ}) significantly contributes to the integral,
only when the following two relations hold concurrently,
\begin{subequations}\label{WiWj_Nonzero}
  \begin{align}
    &H_{\rm cl}\left(\bm{p}-\frac{\bm{p}'}{2},\bm{q}-\frac{\bm{q}'}{2}\right)=E_i,\\
    &H_{\rm cl}\left(\bm{p}+\frac{\bm{p}'}{2},\bm{q}+\frac{\bm{q}'}{2}\right)=E_j,
  \end{align}
\end{subequations}
 meanwhile, $J$ remains substantial.

From Eq.(\ref{J_Definition}), one sees that the function $J$ represents the
Fourier transform of certain correlation of $O_{\rm w}(\bm{p},\bm{q})$.
 For operators $O$ of physical interest, usually
 $O_{\rm w}(\bm{p},\bm{q})$ do not oscillate quite rapidly.
 As a consequence, relatively large values of $J$ usually lie within a region
 with relatively small values of $\bm{p}'/\hbar$ and $\bm{q}'/\hbar$.
 Due to the smallness of $\hbar$, this implies that $J$ is significant
 and show a pronounced peak around the point of $\bm{p}'=\bm{q}'=0$.
 According to Eq.(\ref{WiWj_Nonzero}), such a peak implies smallness of $\abs{E_i-E_j}$.
 In other words, the matrix of $\overline{\abs{\bra{E_i}O\ket{E_j}}^2}$ should
 show a banded structure around its diagonal elements.
 Indeed, this is a phenomenon that has already been observed in many numerical simulations
 (see, e.g., Ref.\cite{YWW2022PS,WangDingZu_2024,Vidmar_2019,Rigol_2017})
 and here an analytical explanation is supplied.

\paragraph*{{Estimate of Bandwidth}}
We proceed to give an estimate to the bandwidth
of the matrix of $\bra{E_i}O\ket{E_j}$, which is to be denoted by $w_b$.
 That is,  for $|E_i - E_j| <w_b$, one has
 $\overline{\abs{\bra{E_i}O\ket{E_j}}^2}\geqslant\epsilon \left(\max{\overline{\abs{\bra{E_i}O\ket{E_j}}^2}} \right)$,
 where $\epsilon$ is a (small) positive number as a measure for precision.
 Substituting Eq.(\ref{Wigner_Semiclassical}) into Eq.(\ref{Oij_WignerJ}),
one sees that
$\overline{\abs{\bra{E_i}O\ket{E_j}}^2}$ is the average value of $J$ over the two energy
surfaces of $H_{\rm cl}=E_i$ and $H_{\rm cl}=E_j$ in the sense of Eq.(\ref{WiWj_Nonzero}), i.e.,
\begin{equation}
  \overline{\abs{\bra{E_i}O\ket{E_j}}^2}\approx\langle J\rangle_{E_i, E_j}.
\end{equation}
Therefore, $w_b$ may be estimated by the requirement of $\langle J\rangle_{E_i, E_j} \geqslant\epsilon \langle J_{\rm max}\rangle_{E_i, E_j}$,
where $J_{\rm max}$ is the maximum value of $J$ over all $(\bm{p}',\bm{q}')$.

To go on, we note that, when discussing the function of $J$ defined in Eq.(\ref{J_Definition}),
it should be convenient to discuss dependence of $J$ on $\bm{X}'=(\bm{X}'_1,\bm{X}'_2)
=(\bm{p}',\bm{q}')/\hbar$. Through this coordinate transformation, $J$'s dependence on
$O_{\rm w}$ and $\hbar$ is separated and can be written as $J \approx (2\pi\hbar)^{-d}
\ww J(\bm{p},\bm{q},\hbar\bm{p}',\hbar\bm{q}')$, where
\begin{equation}\label{J_Definition_X}
  \begin{aligned}
   \ww J&\left(\bm{p},\bm{q},\bm{X_1}', \bm{X_2}'\right)
    := \int d\widetilde{\bm{p}} d\widetilde{\bm{q}}
    \ e^{-i\widetilde{\bm{q}}\cdot\bm{X}'_1}e^{-i\widetilde{\bm{p}}\cdot\bm{X}'_2}\\
    &\times O_{\rm cl}\left(\bm{p}+\frac{1}{2}\widetilde{\bm{p}},\bm{q}-\frac{1}{2}\widetilde{\bm{q}}\right)
    O_{\rm cl}\left(\bm{p}-\frac{1}{2}\widetilde{\bm{p}},\bm{q}+\frac{1}{2}\widetilde{\bm{q}}\right),
  \end{aligned}
\end{equation}
which does not contain $\hbar$.

For a fixed point of $(\bm{p},\bm{q})$, we use $\mathcal{R}^{\widetilde{J}}_\epsilon$ to
denote that region in the space of $\bm{X}' =(\bm{X}'_1,\bm{X}'_2)$, within which
$\widetilde{J}\geqslant\epsilon\widetilde{J}_{\rm max}$, where $\widetilde{J}_{\rm max}$ is the maximum
value of $\widetilde{J}$ over all $\bm{X}'$. Then, for any fixed $(\bm{p},\bm{q})$, to
find the maximum $\abs{E_i-E_j}$ holding $J\geqslant\epsilon J_{\rm max}$ is equivalent
to finding the maximum value of $\abs{E_i-E_j}$ when $\bm{X}'$ runs through the region
$\mathcal{R}^{\widetilde{J}}_\epsilon$.

As previously discussed, for the points $\bm{X}'$ located within $\mathcal{R}^{\widetilde{J}}_\epsilon$,
$(\bm{p}',\bm{q}')=\hbar\bm{X}'$ are extremely close to $\bm{0}$.
Under such circumstances, according to Eq.(\ref{WiWj_Nonzero}), one has
\begin{equation}\label{E_Linear}
  \abs{E_i-E_j}
  =\hbar\abs{\bm{X}'\cdot\bm{\nabla }H_{\rm cl}},
\end{equation}
where $\bm{\nabla }H_{\rm cl}$ represents the gradient of $H_{\rm cl}(\bm{p},\bm{q})$ in
phase space,
\begin{equation}\label{H_Gradient}
  \bm{\nabla }H_{\rm cl}
  =\left(\frac{\partial H_{\rm cl}}{\partial p_1},\cdots,\frac{\partial H_{\rm cl}}{\partial p_d},
  \frac{\partial H_{\rm cl}}{\partial q_1}\cdots\frac{\partial H_{\rm cl}}{\partial q_d}\right).
\end{equation}

\color{black}

\begin{figure}[!t]
  \centering
  \includegraphics[width=1\linewidth]{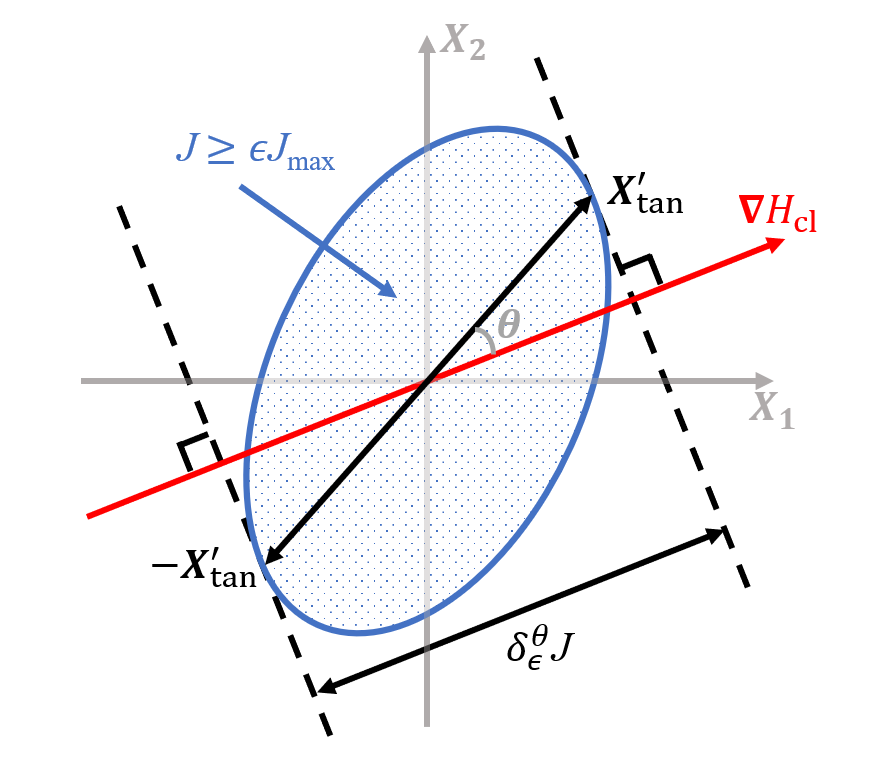}
  \caption{An illustration for the width of a region $\mathcal{R}^{\widetilde{J}}_\epsilon$ along $\bm{\nabla} H_{\rm cl}$.
  }
  \label{J_Width}
\end{figure}

Since $\hbar\abs{\bm{X}'\cdot\bm{\nabla }H_{\rm cl}}$ is a linear function of $\bm{X}'$,
it evidently reaches its maximum at the point $\bm{X}'_{\rm tan}$ —the tangent point
between the region $\mathcal{R}^{\widetilde{J}}_\epsilon$ and a plane perpendicular to
$\bm{\nabla} H_{\rm cl}$ (as illustrated in Fig.\ref{J_Width}). Denoting the angle between
$\bm{X}'_{\rm tan}$ and $\bm{\nabla} H_{\rm cl}$ by $\theta$, according to the
rules of vector multiplication, one finds that $\hbar\abs{\bm{X}'_{\rm tan}\cdot\bm{\nabla }H_{\rm cl}}
=\hbar\abs{\bm{X}'_{\rm tan}}\cdot\abs{\cos{\theta}}\cdot\abs{\bm{\nabla }H_{\rm cl}}$.

 When $\bm{X}'$ running over the region $\mathcal{R}^{\widetilde{J}}_\epsilon$,
 the maximum value of $\abs{E_i-E_j}$
 can be written as $\abs{E_i-E_j}=\hbar\abs{\bm{X}'_{\rm tan}\cdot\bm{\nabla }H_{\rm cl}}
=\frac{1}{2}\hbar\abs{\bm{\nabla }H_{\rm cl}}\delta_\epsilon^\theta \widetilde{J}$,
 where $\delta_\epsilon^\theta \widetilde{J}:=2\abs{\bm{X}'_{\rm tan}} \abs{\cos{\theta}}$.
 Since $\widetilde{J}\left(\bm{p},\bm{q},-\bm{X}'\right)=\widetilde{J}\left(\bm{p},\bm{q},\bm{X}'\right)$,
 which follows easily from the definition in Eq.(\ref{J_Definition_X}),
 $-\bm{X}'_{\rm tan}$ is also a tangent point for the region $\mathcal{R}^{\widetilde{J}}_\epsilon$
and a plane perpendicular to $\bm{\nabla} H_{\rm cl}$.
 Hence, $\delta_\epsilon^\theta \widetilde{J}$ represents
the width of region $\mathcal{R}^{\widetilde{J}}_\epsilon$ along $\bm{\nabla} H_{\rm cl}$
 (see Fig.\ref{J_Width}).

Thus far, we have established that, at a fixed $(\bm{p},\bm{q})$, the maximum
value of $\abs{E_i-E_j}$
that keeps $J\geqslant\epsilon J_{\rm max}$ is $\frac{1}{2}\hbar\abs{\bm{\nabla }H_{\rm cl}}\delta_\epsilon^\theta \widetilde{J}$.
Considering contributions from all $(\bm{p},\bm{q})$ integrated in Eq.(\ref{Oij_WignerJ}),
 one sees that the bandwidth of the matrix of $\bra{E_i}O\ket{E_j}$ should be equal to the average value of
$\frac{1}{2}\hbar\abs{\bm{\nabla }H_{\rm cl}}\delta_\epsilon^\theta \widetilde{J}$ over all $(\bm{p},\bm{q})$
within the neighborhood of the energy surface of $H_{\rm cl}=(E_i+E_j)/2$,
 written as $\frac{1}{2}\hbar \langle\abs{\bm{\nabla }H_{\rm cl}}\delta_\epsilon^\theta \widetilde{J}\rangle$.

Note that $\delta_\epsilon^\theta \widetilde{J}$ fluctuates randomly around $\delta_\epsilon \widetilde{J}$,
where $\delta_\epsilon \widetilde{J}$ is the mean value of $\delta_\epsilon^\theta \widetilde{J}$
averaged over all $\theta$,\footnote{For special cases in which the distribution of $\theta$
is uneven, one may just go back to $\abs{E_i-E_j}\approx\hbar\abs{\bm{X}'_{\rm tan}\cdot\bm{\nabla }H_{\rm cl}}$
for application.}
 which in fact represents the average width of $\mathcal{R}^{\widetilde{J}}_\epsilon$
 over all directions.
 Hence, in $\frac{1}{2}\hbar \langle\abs{\bm{\nabla }H_{\rm cl}}\delta_\epsilon^\theta \widetilde{J}\rangle$,
 one may replace $\delta_\epsilon^\theta \widetilde{J}$ by $\delta_\epsilon \widetilde{J}$.
 Moreover, since $J$ is independent of the Hamiltonian,
 average of $\abs{\bm{\nabla }H_{\rm cl}}$ and that of $\delta_\epsilon \widetilde{J}$ may be taken separately,
 when carrying out the average over $(\bm{p},\bm{q})$.
 Thus, we reach the conclusion that the matrix of ${{\bra{E_i}O\ket{E_j}}}$ has a band
structure with a width
\begin{align}\label{BandWidth}
  w_b \simeq \frac{1}{2}\hbar \langle\abs{\bm{\nabla }H_{\rm cl}}\rangle\langle\delta_\epsilon \widetilde{J}\rangle.
\end{align}

 The above obtained width depends mainly on three quantities: $\hbar$, a Hamiltonian-dependent part
$\abs{\bm{\nabla }H_{\rm cl}}$, and an observable-dependent part $\delta_\epsilon \widetilde{J}$.
 Whenever being interested in a fixed observable,
 the value of $\langle\delta_\epsilon \widetilde{J}\rangle$ is fixed and, then,
 the band width $w_b$ is mainly dominated by $\abs{\bm{\nabla }H_{\rm cl}}$.
 Note that, taking $\epsilon=\frac{1}{2}$,
 $w_b$ gives the half width of  $\overline{\abs{\bra{E_i}O\ket{E_j}}^2}$ as a function of $(E_i-E_j)$.

 As an importance case in the application of ETH,
 suppose that the total system is divided into two subsystems, $A$ and $B$,
 with $O$ as observable of $A$ indicated as $O^A$.
 From Eq.(\ref{J_Definition}), it is direct to find that
 $J=(2\pi\hbar)^{d_B}\delta^{d_B}(\bm{p}'_B)\delta^{d_B}(\bm{q}'_B)J_A$,
 where $d_B$ is the number of degrees of freedom of the system $B$ and
 $J_A$ is defined in the same way as $J$ in Eq.(\ref{J_Definition}), but for coordinates of $A$.
 This gives that
\begin{align}\label{wb-OA}
 w_b \simeq \frac{1}{2}\hbar\langle\abs{\bm{\nabla}_A H_{\rm cl}}\rangle\langle\delta_\epsilon \widetilde{J}_A\rangle.
\end{align}
 One interesting prediction is that $w_b$ is independent of the size of the system $B$,
 a phenomenon that has been observed in numerical simulations
 (see, e.g., Refs.\cite{YWW2022PS,Gemmer_2020}).

\paragraph*{$\rho_{\rm dos}$-scaling.}
 Equation (\ref{Oij_WignerJ}) predicts the $\rho_{\rm dos}$-scaling behavior of
 the offdiagonal elements $\bra{E_i}O\ket{E_j}$ in the ETH ansatz,
 in the case that $\hbar$ is small and  $O_{\rm cl}$ is a relatively simple function of $(\bm{p},\bm{q})$.
 In fact, according to Eq.(\ref{Osc-Ocl}), smallness of $\hbar$ implies that
 the functions of $O_{\rm w}$ on the right-hand side (rhs) of Eq.(\ref{J_Definition})
 may be approximated by $O_{\rm cl}$.
 Then, for relatively simple functions of $O_{\rm cl}$,
 due to the terms of $e^{-i\widetilde{\bm{p}}\cdot\bm{q}'/\hbar}e^{-i\bm{p}'\cdot\widetilde{\bm{q}}/\hbar}$,
 the $2d$-dimensional integration on the rhs  of Eq.(\ref{J_Definition})
 usually generates a factor scaling as $\hbar^{2d}$, which we write as $(2\pi\hbar)^{2d}$
(see Sec.\ref{J_Scaling} of Supplementary Materials).
Finally, cancelling the prefactor of $(2\pi\hbar)^{-d}$ on the rhs of Eq.(\ref{J_Definition}),
$J$ should contain a factor of $(2\pi\hbar)^d$, and similar for $\overline{\abs{\bra{E_i}O\ket{E_j}}^2}$.
Since $\rho_{\rm dos}  \sim(2\pi\hbar)^{-d}$,
one finds that $\overline{\abs{\bra{E_i}O\ket{E_j}}^2} \sim 1/\rho_{\rm dos}$,
 which predicts that $\bra{E_i}O\ket{E_j} \sim 1/\sqrt{\rho_{\rm dos}}$.

\paragraph*{{Numerical Simulations}}
We have numerically verified results presented above in  the three-orbital
LMG model \cite{Lipkin_1965,WIC98}. This model is composed of
$\Omega$ particles, occupying three orbitals each with $\Omega$-degeneracy.
In this model, $\hbar_{\rm eff}=1/\Omega$ plays the role of an effective Planck's constant.
 The system consists of two degrees of freedom, with $A$
 for a first one and $B$ for the second.
 We study a parameter regime of the model, in which it is a quantum chaotic system
 with a Hamiltonian written as $H= H_A + H_B +V$, where $V$ represents the coupling.
 The operator $O^A$ is $O^A=b_{1}^{\dagger}b_{1}/\Omega$, where $b_{1}:=\sqrt{\frac{\Omega}{2}}(\hat{q}_{1}+i\hat{p}_{1})$.
 It is easy to find that
 $O_{\rm w}^A=\frac{1}{2}\left(p^2_1+q^2_1\right)-\frac{1}{2}\hbar_{\rm eff}$ \cite{WangXiao_2024}.
(See Sec.\ref{Sec_LMG_Model} of the Supplementary Materials for details of the model.)

\begin{figure}[!t]
  \centering
  \includegraphics[width=1\linewidth]{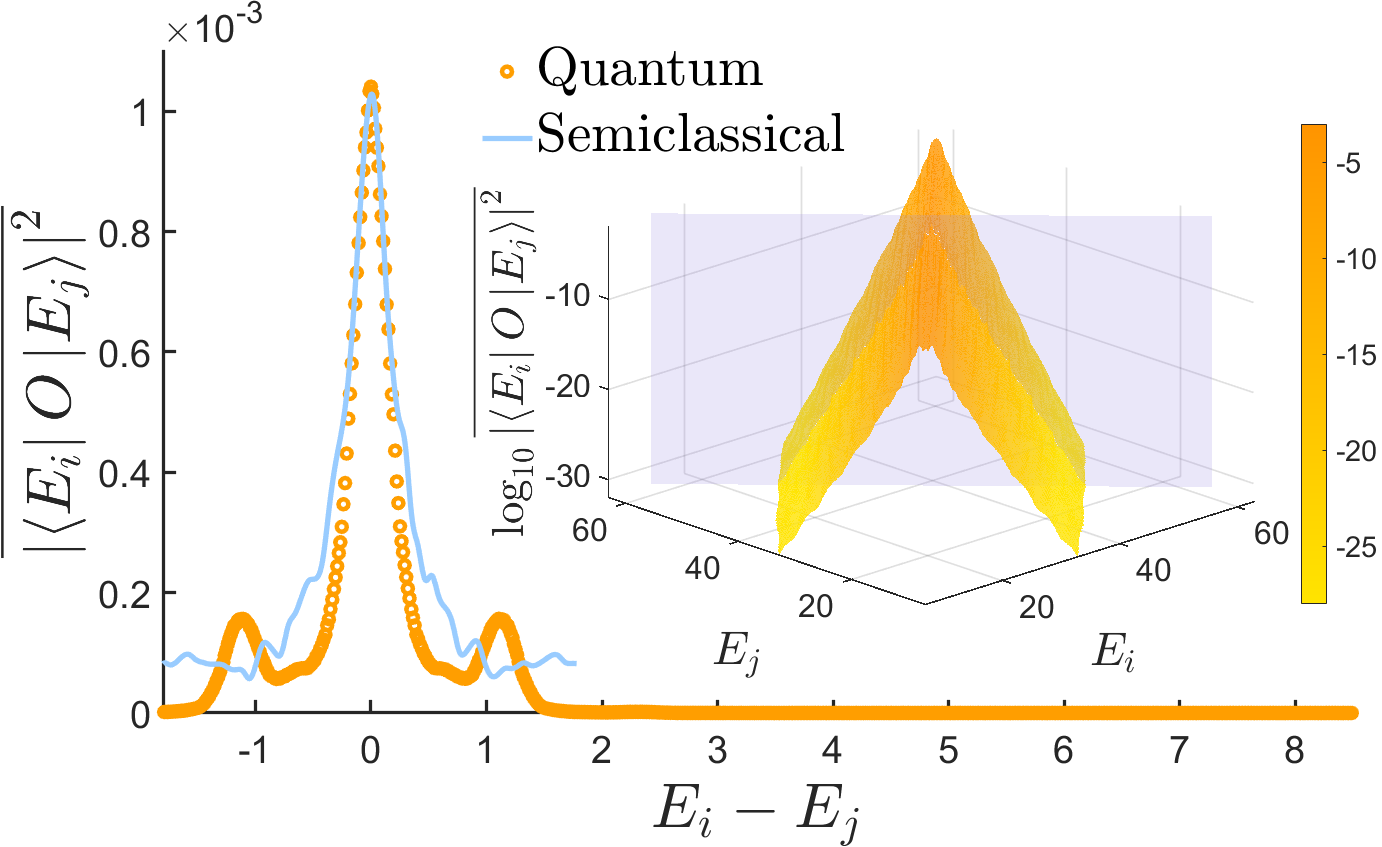}
  \caption{
  Variation of $\overline{\abs{\bra{E_i}O\ket{E_j}}^2}$ (open circles) with $(E_i -E_j)$,
  for a fixed value of $\frac{1}{2}(E_i+E_j)$ at the center of the energy spectrum.
 The solid line (blue) represents the semiclassical prediction by Eqs.(\ref{Oij_WignerJ}) and Eq.(\ref{Wigner_Semiclassical}).
  The 3D inset shows the overall shape of $\overline{\abs{\bra{E_i}O\ket{E_j}}^2}$ in a logarithmic scale,
  with the position of the main plot indicated by a (purple) plane.
  }
  \label{f_Compare}
\end{figure}

 The shape of $\overline{\abs{\bra{E_i}O\ket{E_j}}^2}$ is plotted in Fig.\ref{f_Compare},
 with average taken over narrow energy shells around $E_i$ and $E_j$ for the purpose of suppressing fluctuations
 (approximately 25 levels in each shell).
 The figure shows that the matrix of $\overline{\abs{\bra{E_i}O\ket{E_j}}^2}$ has a band structure
 with respect to $(E_i - E_j)$.
 And, the semiclassical expression of Eq.(\ref{Oij_WignerJ}),
 with the Wigner function approximated by Eq.(\ref{Wigner_Semiclassical}),
 works well for the central peak (main body) of $\overline{\abs{\bra{E_i}O\ket{E_j}}^2}$.

\begin{figure}[!t]
  \centering
  \includegraphics[width=1\linewidth]{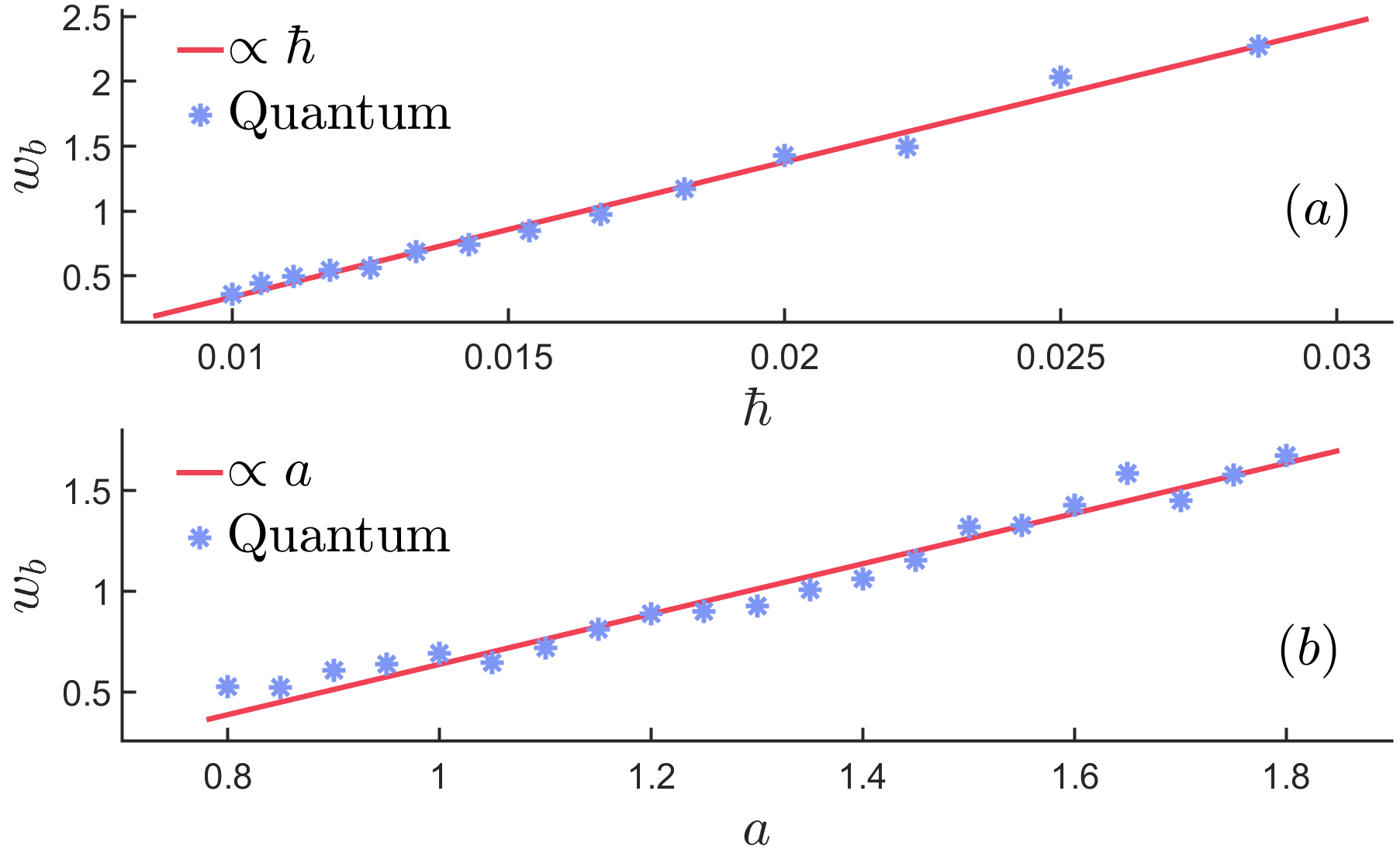}
  \caption{ (a)
  Dependence of $w_b$, the half width of $\overline{\abs{\bra{E_i}O\ket{E_j}}^2}$ (stars in blue),
  on the effective Planck constant $\hbar_{\rm eff}$.
  The line (red) shows a linear fit to the stars.
  (b) Dependence of $w_b$ on the  parameter $a\propto \bm{\nabla}_A H$.
  }
  \label{ETH_fWidth}
\end{figure}

\begin{figure}[!t]
  \centering
  \includegraphics[width=1\linewidth]{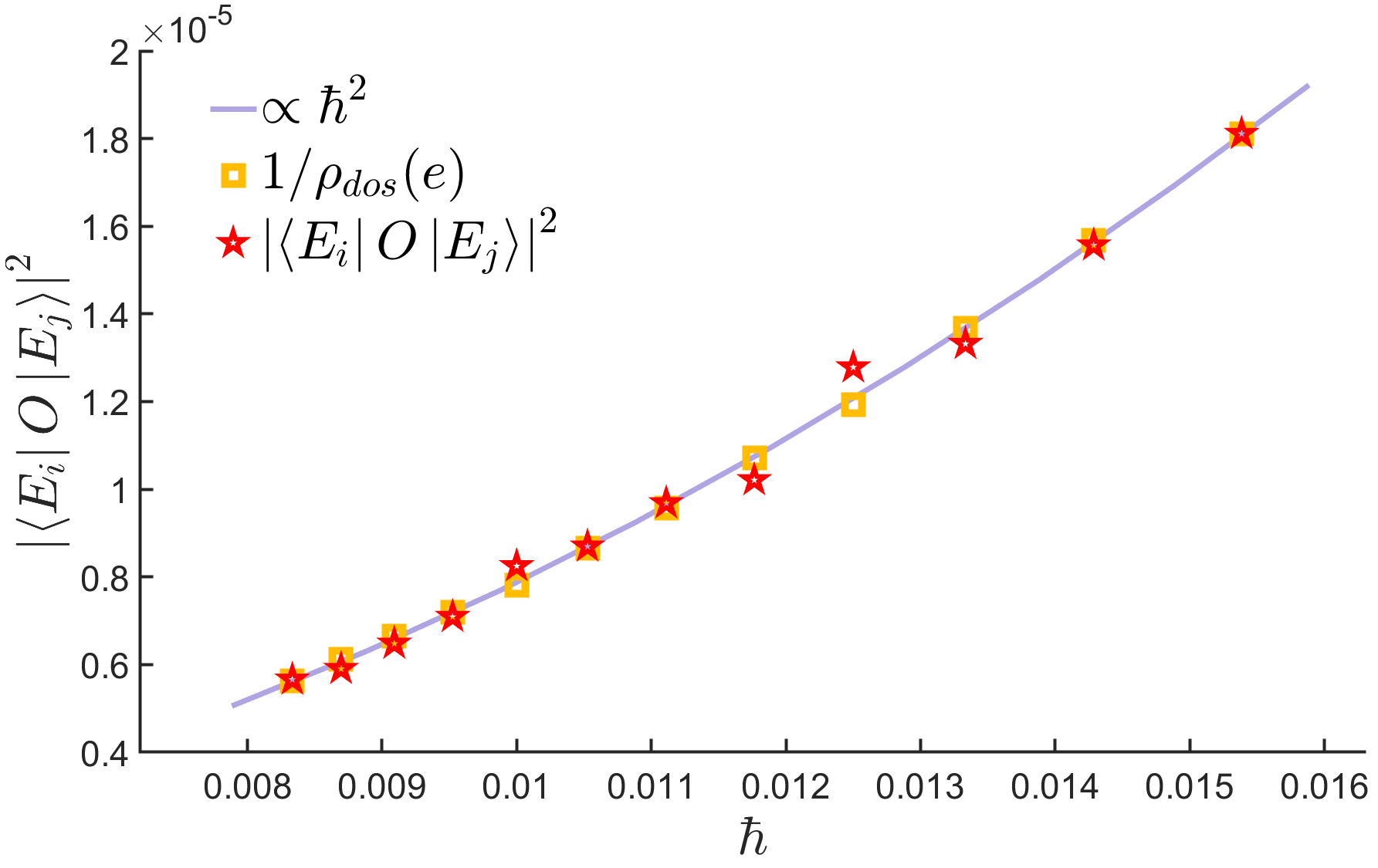}
  \caption{Variation of the average magnitude of $\abs{\bra{E_i}O\ket{E_j}}^2$ (red stars)
  with $\hbar$. The average was taken over those points of $(E_i, E_j)$
  at which $e=(E_i+E_j)/2$ is located
  at the center of the spectrum. The yellow squares represent the $1/\rho_{\rm dos}$ at $e$. In
  this model, $d=2$ and, hence, $\rho_{\rm dos}\propto \hbar^{-2}$.\color{black}
  }
  \label{ETH_fHeight}
\end{figure}

Figure \ref{ETH_fWidth}(a) depicts variation of the halfwidth of $\overline{\abs{\bra{E_i}O\ket{E_j}}^2}$
 versus $\hbar$, showing the $\hbar$ behavior predicted in Eq.(\ref{wb-OA}).
 In order to test the prediction of $w_b \sim |\bm{\nabla}_A H_{\rm cl}|$,
 we introduce a parameter $a$ to $H$, $H(a)=a H_A + aV + H_B$,
 such that $\bm{\nabla}_A H_{\rm cl} \propto a $.
 Fig.\ref{ETH_fWidth}(b) shows that $w_b$ increases approximately linearly with $a$,
 with some fluctuations which possibly due to the weak distortion of the energy surface resulting from $a$.
 In addition, we have also numerically checked the prediction of
 $\overline{\abs{\bra{E_i}O\ket{E_j}}^2} \propto \hbar_{\rm eff}^{-2}$,
 which implies that $\overline{\abs{\bra{E_i}O\ket{E_j}}^2}\propto 1/\rho_{\rm dos}$
 (Fig.\ref{ETH_fHeight}).

\paragraph*{Thermalization Time}
 The expression of $w_b$ in Eq. (\ref{BandWidth}) may be used to
 give an estimate to the thermalization time $\tau$, which is generally expected to
 scale as $\tau \sim \hbar/w_b$
\cite{Jiaozi_2022,Schiulaz_2019,Serbyn_2017,Prosen_Vidmar_2020,Prosen_2021}.
 It predicts $\tau \sim \langle|\bm{\nabla} H_{\rm cl}|\rangle^{-1}$,
 in agreement with the expectation that
 smaller $\langle|\bm{\nabla} H_{\rm cl}|\rangle$ implying slower thermalization.
 Indeed, according to the classical Hamilton's equations,
 one has $   \frac{\partial H_{\rm cl}}{\partial p_\mu}=\dot{q}_\mu \ \& \
 \frac{\partial H_{\rm cl}}{\partial q_\mu}=-\dot{p}_\mu=-F_\mu$,
 with small $\langle|\bm{\nabla} H_{\rm cl}|\rangle$ implying slow diffusion
 and energy transfer.
 Moreover, for a total system satisfying ETH with the division of $A+B$,
 Eq.(\ref{wb-OA}) predicts that the thermalization time of the subsystem $A$
 dependents on neither the size nor the detailed dynamics of its environment $B$.

\paragraph*{Conclusions and Discussions}
 A semiclassical expression for the offdiagonal function
 in ETH has been derived. The expression predicts a banded structure of the matrix of
 ${\bra{E_i}O\ket{E_j}}$, which is a strong constraint on the matrix behavior.
 The analytically derived bandwidth $w_b$ has an interesting feature:
 as a product of $\hbar$, the absolute phase-space gradient of the classical Hamiltonian,
 and a property of the studied operator $O$  [Eq.(\ref{BandWidth})].
 This predicts that the thermalization time, which is usually expected to scale as $w^{-1}_b$,
 may be proportional to $|\bm{\nabla }H_{\rm cl}|^{-1}$.
 In addition, we elucidate the origin of the $\rho_{\rm dos}^{-1/2}$-scaling behavior
 of offdiagonal elements in ETH.

\paragraph*{Acknowledgments}
The authors are grateful to Dr. Jiaozi Wang for valuable discussions and suggestions. 
This work was partially supported by the Natural Science Foundation of China under Grant
 Nos. 12175222, 92565306, and 11775210.

\appendix

\section{Derivation of E\lowercase{q}.(\ref{Oij_WignerJ})}\label{Derivation_Oij_WignerJ}

In this section, we derive Eq.(\ref{Oij_WignerJ}) in the main text.
 Let us insert identity operators into the definition of $\abs{\bra{E_i}O\ket{E_j}}^2$,  getting that
\begin{equation}\label{Oij_Derivation_1}
    \begin{aligned}
        &\abs{\bra{E_i}O\ket{E_j}}^2
        = \int d\bm{q}_1d\bm{q}_2\braket{E_i}{\bm{q}_1}\bra{\bm{q}_1}O\ket{\bm{q}_2}\braket{\bm{q}_2}{E_j}\\
        &\qquad\qquad\times\int d\bm{q}_3d\bm{q}_4\braket{E_j}{\bm{q}_3}\bra{\bm{q}_3}O\ket{\bm{q}_4}\braket{\bm{q}_4}{E_i}.
    \end{aligned}
\end{equation}
Performing inverse Fourier transform on the Wigner function $W_i$ (Eq.(\ref{WignerFunction_Definition})) and
Weyl symbol $O_{\rm w}(\bm{p},\bm{q})$ (Eq.(\ref{Osc_Definition})), we find that
\begin{subequations}\label{WiOsc2Elements}
    \begin{align}
          &\braket{\bm{q}_1}{E_i}
          \braket{E_i}{\bm{q}_2}=\int d\bm{p}
  \ W_i\left(\bm{p},\frac{\bm{q}_1+\bm{q}_2}{2}\right)
  e^{-i\bm{p}\cdot(\bm{q}_2-\bm{q}_1)/\hbar},\\
&\bra{\bm{q}_1}O(\hat{\bm{p}},\hat{\bm{q}})\ket{\bm{q}_2}\notag\\
  &\qquad\quad=\frac{1}{(2\pi\hbar)^d}\int d\bm{p}\ O_{\rm w}\left(\bm{p},\frac{\bm{q}_1+\bm{q}_2}{2}\right)e^{i \bm{p}\cdot(\bm{q}_1-\bm{q}_2)/\hbar}
  .
    \end{align}
\end{subequations}
Substituting Eq.(\ref{WiOsc2Elements}) into Eq.(\ref{Oij_Derivation_1}), we get that
\begin{equation}\label{Oij_Derivation_2}
    \begin{aligned}
        &\abs{\bra{E_i}O\ket{E_j}}^2
        =\int d\bm{q}_1d\bm{q}_2d\bm{q}_3d\bm{q}_4
        \int d\bm{p}_1d\bm{p}_2d\bm{p}_3d\bm{p}_4\\
        &\quad\frac{1}{(2\pi\hbar)^{2d}}
        W_i\left(\bm{p}_1,\frac{\bm{q}_1+\bm{q}_4}{2}\right)
        W_j\left(\bm{p}_2,\frac{\bm{q}_2+\bm{q}_3}{2}\right)
  \\
        &\times
        O_{\rm w}\left(\bm{p}_3,\frac{\bm{q}_1+\bm{q}_2}{2}\right)
        O_{\rm w}\left(\bm{p}_4,\frac{\bm{q}_3+\bm{q}_4}{2}\right)\\
        &\times
        e^{-i\bm{p}_1\cdot(\bm{q}_1-\bm{q}_4)/\hbar}
  e^{-i\bm{p}_2\cdot(\bm{q}_3-\bm{q}_2)/\hbar}\\
  &\times
  e^{i \bm{p}_3\cdot(\bm{q}_1-\bm{q}_2)/\hbar}
        e^{i \bm{p}_4\cdot(\bm{q}_3-\bm{q}_4)/\hbar}.
    \end{aligned}
\end{equation}

Let us perform the following variable substitution,
\begin{equation}
    \begin{aligned}
        &\bm{p}=\frac{\bm{p}_1+\bm{p}_2}{2},
        &\bm{p}'=\bm{p}_2-\bm{p}_1,
        \qquad\qquad\quad\\
        &\widetilde{\bm{p}}=\bm{p}_3-\bm{p}_4,
        \qquad&\widetilde{\bm{p}'}=\frac{\bm{p}_3+\bm{p}_4}{2},
        \qquad\qquad\quad\\
        &\bm{q}=\frac{\bm{q}_1+\bm{q}_2+\bm{q}_3+\bm{q}_4}{4},
        &\bm{q}'=\frac{\bm{q}_2+\bm{q}_3}{2}-\frac{\bm{q}_1+\bm{q}_4}{2},
        \\
        &\widetilde{\bm{q}}=\frac{\bm{q}_3+\bm{q}_4}{2}-\frac{\bm{q}_1+\bm{q}_2}{2},
        &\widetilde{\bm{q}'}=\frac{\bm{q}_1+\bm{q}_3}{2}-\frac{\bm{q}_2+\bm{q}_4}{2}.
    \end{aligned}
\end{equation}
 This gives that
\begin{equation}\label{Oij_Derivation_3}
    \begin{aligned}
        &\abs{\bra{E_i}O\ket{E_j}}^2
        =2^d\int d\bm{p}d\bm{p}'d\widetilde{\bm{p}}d\widetilde{\bm{p}'}
        \int d\bm{q}d\bm{q}'d\widetilde{\bm{q}}d\widetilde{\bm{q}'}\\
        &\frac{1}{(2\pi\hbar)^{2d}}
        W_i\left(\bm{p}-\frac{\bm{p}'}{2},\bm{q}-\frac{\bm{q}'}{2}\right)
        W_j\left(\bm{p}+\frac{\bm{p}'}{2},\bm{q}+\frac{\bm{q}'}{2}\right)
  \\
        &\times
        O_{\rm w}\left(\widetilde{\bm{p}'}+\frac{\widetilde{\bm{p}}}{2},\bm{q}-\frac{\widetilde{\bm{q}}}{2}\right)
        O_{\rm w}\left(\widetilde{\bm{p}'}-\frac{\widetilde{\bm{p}}}{2},\bm{q}+\frac{\widetilde{\bm{q}}}{2}\right)\\
        &\times
e^{-i\bm{p}'\cdot\widetilde{\bm{q}}/\hbar}
  e^{-i \widetilde{\bm{p}}\cdot\bm{q}'/\hbar}
    e^{2 i (\widetilde{\bm{p}'}-\bm{p})\cdot\widetilde{\bm{q}'}/\hbar}.
        \\
    \end{aligned}
\end{equation}
 Making use of the following property,
\begin{equation}
    \int d\widetilde{\bm{q}'} e^{2 i (\widetilde{\bm{p}'}-\bm{p})\cdot\widetilde{\bm{q}'}/\hbar}=\left(\pi\hbar\right)^d\delta(\widetilde{\bm{p}'}-\bm{p}),
\end{equation}
 and  carrying out the integration over $\widetilde{\bm{q}'}$ and $\widetilde{\bm{p}'}$, finally we get
\begin{equation}\label{Oij_Derivation_4}
    \begin{aligned}
        &\abs{\bra{E_i}O\ket{E_j}}^2
        =\frac{1}{(2\pi\hbar)^d}
        \int d\bm{p}d\bm{p}'d\widetilde{\bm{p}}
        \int d\bm{q}d\bm{q}'d\widetilde{\bm{q}}\\
        &\qquad
        W_i\left(\bm{p}-\frac{\bm{p}'}{2},\bm{q}-\frac{\bm{q}'}{2}\right)
        W_j\left(\bm{p}+\frac{\bm{p}'}{2},\bm{q}+\frac{\bm{q}'}{2}\right)
  \\
        &\quad\times
        O_{\rm w}\left(\bm{p}+\frac{\widetilde{\bm{p}}}{2},\bm{q}-\frac{\widetilde{\bm{q}}}{2}\right)
        O_{\rm w}\left(\bm{p}-\frac{\widetilde{\bm{p}}}{2},\bm{q}+\frac{\widetilde{\bm{q}}}{2}\right)\\
        &\quad\times
e^{-i\bm{p}'\cdot\widetilde{\bm{q}}/\hbar}
  e^{-i \widetilde{\bm{p}}\cdot\bm{q}'/\hbar}.
        \\
    \end{aligned}
\end{equation}
 This finishes proof of Eq.(\ref{Oij_WignerJ}).

\section{Weyl Symbol of Observable}\label{Osc_Introduction}
\setcounter{equation}{0}

In this section, we briefly recall properties of the Weyl Symbol $O_{\rm w}(\bm{p},\bm{q})$
discussed in Ref.\cite{WangXiao_2024},
which are closely related to the concept of \textit{Weyl order}.
For brevity, we discuss the case of a one-dimensional configuration space. The
results high-dimensional in cases are similar.

An operator is said to be Weyl-ordered, indicated as
$A^{\rm Weyl}$, if it has the following binomial form \cite{Peskin_2019},
\begin{equation}\label{WeylOrder}
 A^{\rm Weyl}(\hat{p},\hat{q})=\sum_{n=0}^\infty \frac{1}{2^n}
 \sum_{k=0}^n \binom{n}{k}\hat{q}^{n-k}g_n(\hat{p})\hat{q}^k,
\end{equation}
where $g_n(\hat{p})$ is some function of $\hat{p}$, which is $k$-independent while may
be $n$-dependent, and $\binom{n}{k}$ is the combination number,
\begin{equation}
 \binom{n}{k} =\frac{n!}{k!(n-k)!}.
\end{equation}
For such an operator, using the definition in Eq.(\ref{Osc_Definition}) of the main text, it's easy to
prove by direct calculations that the corresponding Weyl symbol has the following form,
\begin{equation}\label{Weyl_Sc}
  A_{\rm w}(p,q)=\sum_{n=0}^\infty q^n g_n(p).
\end{equation}
Note that this form is exactly the same as the expression gotten by directly
replacing the operators $(\hat{p},\hat{q})$ in $A^{\rm Weyl}(\hat{p},\hat{q})$
by the classical coordinates $(p,q)$.

Although an operator $O(\hat{p},\hat{q})$,
 as a general polynomial function of $\hat{p}$ and $\hat{q}$,
 usually doesn't have the Weyl-ordered form defined in Eq.(\ref{WeylOrder}),
 it may be written as a sum of Weyl-ordered operators, by
 commutating $\hat{p}$ and $\hat{q}$
 \cite{WangXiao_2024,Peskin_2019}.
 Among the latter, there is only one term, denoted by $O^{\rm Weyl}(\hat{p},\hat{q})$,
 which contains the same power of $\hat{p}$ and $\hat{q}$,
 meanwhile, others contains at least only $\hbar$ as a consequence of
 commutating $\hat{p}$ and $\hat{q}$.
 For example, for $O=\hat{q}\hat{p}\hat{q}\hat{p}$, one finds that
\begin{align}\label{}
 O = O^{\rm Weyl}  +\frac{3}{8}i\hbar\left(\hat{q}\hat{p}+\hat{p}\hat{q}\right)+\frac{1}{8}\hbar^2,
\end{align}
where
\begin{equation}
  \begin{aligned}
 & O^{\rm Weyl} =\frac{1}{4}\left(\hat{q}^2\hat{p}^2+2\hat{q}\hat{p}^2\hat{q}+\hat{p}^2\hat{q}^2\right).
  \end{aligned}
\end{equation}

 To summarize, $O(\hat{p},\hat{q})$ is written as a sum of
 Weyl-ordered operators, with the first one as $O^{\rm Weyl}(\hat{p},\hat{q})$
 and the rest containing coefficient $\hbar^m$ of $m \geqslant 1$.
 This implies the following difference between $O_{\rm w}(p,q)$ and $O_{\rm cl}(p,q)$,
\begin{equation}\label{Osc_Ocl_Error}
  O_{\rm w}(p,q) = O_{\rm cl}(p,q) + {\rm O}(\hbar^m) \quad (m \geqslant 1).
\end{equation}
This equation also holds in high-dimensional cases \cite{WangXiao_2024}.

\section{The LMG Model}\label{Sec_LMG_Model}
\setcounter{equation}{0}

In this section, we give a brief discussion on the Lipkin-Meshkov-Glick (LMG)
model, which is used in the numerical simulations of the main text.

The employed model is a three-orbital LMG model \cite{Lipkin_1965}.
 This model is composed of $\Omega$ particles,
 occupying three energy orbitals labeled by $r=0,1,2$, each with
 $\Omega$-degeneracy.
 Here, we are interested in the collective motion of this model, for which the dimension of the
 Hilbert space is $\frac 12 (\Omega+1)(\Omega +2)$  \cite{WIC98}.
 We use $\epsilon_{r}$ to denote the energy of the $r$-th level
 and, for brevity, set $\epsilon_{0}=0$.

 The Hamiltonian of the model is written as
\begin{equation}\label{H_LMG}
H=H_0+\lambda V,
\end{equation}
 where $H_0$ indicates the Hamiltonian of an integrable system and $V$ is a perturbation.
 Specifically,
\begin{gather}
 H_{0}=\epsilon_{1}K_{11}+\epsilon_{2}K_{22},  \label{H0}\\
V=\sum_{t=1}^{4}\mu_{t}V^{(t)},
\end{gather}
where
\begin{equation}
  \begin{aligned}
    V^{(1)}=K_{10}K_{10}+K_{01}K_{01},\ V^{(2)}=K_{20}K_{20}+K_{02}K_{02},\\
 V^{(3)}=K_{21}K_{20}+K_{02}K_{12},\ V^{(4)}=K_{12}K_{10}+K_{01}K_{21}.
  \end{aligned}
\end{equation}
Here, the operators $K_{rs}$ are defined by
\begin{equation}
K_{rs}=\sum_{\gamma=1}^{\Omega}a_{r\gamma}^{\dagger}a_{s\gamma},\quad r,s=0,1,2,
\end{equation}
where $a^{\dagger}_{r\gamma}$ and $a_{r\gamma}$ are fermionic creation and annihilation
operators obeying the usual anti-commutation relations.

For symmetric states, the operators $K_{rs}$ can be written
in terms of bosonic creation and annihilation operators $b^\dagger _r$ and $b_r$ \cite{Gong-ou_Xu_1995},
\begin{equation}
K_{rs}=b_{r}^{\dagger}b_{s},\quad K_{r0}=K_{0r}^{\dagger}=b_{r}^{\dagger}
\sqrt{\Omega-b_{1}^{\dagger}b_{1}-b_{2}^{\dagger}b_{2}},
\end{equation}
for $r,s=1,2$.
 Under the transformation,
\begin{equation}
b_{r}^{\dagger}=\sqrt{\frac{\Omega}{2}}(\hat{q}_{r}-i\hat{p}_{r}),\ \ \ b_{r}=\sqrt{\frac{\Omega}{2}}(\hat{q}_{r}+i\hat{p}_{r}),
\end{equation}
for $r=1,2$, it is easy to verify that $q_{r}$ and $p_{s}$ obey the following commutation relation,
\begin{equation}
[\hat{q}_{r},\hat{p}_{s}]=\frac{i}{\Omega}\delta_{rs}.
\end{equation}
 Hence, $1/\Omega$ plays the role of an effective Planck constant,
\begin{equation}
\hbar_{\rm eff}=\frac{1}{\Omega}.
\end{equation}

 It is straightforward to find that the classical counterpart of the model,
 which possesses a two-dimensional configuration space ($d=2$),
 has the following Hamiltonian \cite{WIC98,Meredith88},
\begin{equation}\label{H_LMG_Classical}
H_{\rm cl}(\boldsymbol{p},\boldsymbol{q})=H_0(\boldsymbol{p},\boldsymbol{q})
 +\lambda V(\boldsymbol{p},\boldsymbol{q}),
\end{equation}
where
\begin{gather}\label{LMG_Cl_H0V}
H^{\rm cl}_{0}(\boldsymbol{p},\boldsymbol{q})=\frac{\epsilon'_{1}}{2}(p_{1}^{2}+q_{1}^{2})+\frac{\epsilon'_{2}}{2}(p_{2}^{2}+q_{2}^{2}), \nonumber \\
V_{\rm cl}(\boldsymbol{p},\boldsymbol{q})=\mu'_{1}(q_{1}^{2}-p_{1}^{2})(1-G/2)+\mu'_{2}(q_{2}^{2}-p_{2}^{2})(1-G/2) \nonumber \\
+\frac{\mu'_{3}}{\sqrt{2}}[(q_{2}^{2}-p_{2}^{2})q_{1}-2q_{2}p_{1}p_{2}]\sqrt{1-G/2} \nonumber \\
+\frac{\mu'_{4}}{\sqrt{2}}[(q_{1}^{2}-p_{1}^{2})q_{2}-2q_{1}p_{1}p_{2}]\sqrt{1-G/2},
\end{gather}
with $G=q_{1}^{2}+p_{1}^{2}+q_{2}^{2}+p_{2}^{2}\le2$.
 Here,  the classical parameters are given by
 $\epsilon'_{1}=\epsilon_{1} \Omega, \epsilon'_{2}=\epsilon_{2} \Omega,
 \mu'_{1} = \mu_{1} \Omega^2, \mu'_{2} = \mu_{2} \Omega^2, \mu'_{3}= \mu_{3} \Omega^2$,
and $\mu_{4}'=\mu_{4} \Omega^2$.

In numerical simulations, fixed classical parameters were used, which are $\epsilon'_{1}=44.00,  \epsilon'_{2}= 64.40,
 \mu'_{1}= 18.56, \mu'_{2} = 27.40, \mu'_{3}= 25.28$, $\mu'_{4} = 7.024$,
$\lambda=2$. Under these parameters, different values of the particle number $\Omega$ correspond to a same classical counterpart, which is chaotic.
Properties of the LMG model in the quantum chaotic region have been studied well in previous works (e.g., see Refs.\cite{Meredith88,WIC98,pre18-EF-BC}).

For the purpose of testing the prediction in the main text that $w_b\propto\abs{\bm{\nabla}_A H_{\rm cl}}$,
we further introduce a parameter $a$ into the Hamiltonian:
\begin{equation}
  H=a\cdot\epsilon_{1}K_{11}+\epsilon_{2}K_{22}+a\cdot\lambda V.
\end{equation}
Obviously, in this case,
\begin{equation}
  \bm{\nabla}_A H_{\rm cl}=\left(\frac{\partial H_{\rm cl}}{\partial q'_1},
  \frac{\partial H_{\rm cl}}{\partial p'_1}\right)\propto a.
\end{equation}

 Two remarks: The parameter $a$ is used in the main text only in Fig.\ref{ETH_fWidth}.
 And, in numerical simulations of the main text, the observable $O$ is always taken as
$K_{11}/\Omega$, the corresponding $O_{\rm w}(\bm{p},\bm{q})=\frac{1}{2}\left(p^2_1+q^2_1\right)-\frac{1}{2}\hbar_{\rm eff}$ \cite{WangXiao_2024}.

\section{Examples for $J$ Functions}\label{SupM_Jfun}
\setcounter{equation}{0}

In this section,
we give some examples for the bilinear phase-space convolution function $J$ in
Eq.(\ref{J_Definition}) of the main text,
\begin{equation}
\notag
  \begin{aligned}
    J&\left(\bm{p},\bm{q},\bm{p}',\bm{q}'\right)
    =\frac{1}{(2\pi\hbar)^d}\int d\widetilde{\bm{p}} d\widetilde{\bm{q}}
    \ e^{-i\widetilde{\bm{p}}\cdot\bm{q}'/\hbar}e^{-i\bm{p}'\cdot\widetilde{\bm{q}}/\hbar}\\
    &\times O_{\rm w}\left(\bm{p}+\frac{1}{2}\widetilde{\bm{p}},\bm{q}-\frac{1}{2}\widetilde{\bm{q}}\right)
    O_{\rm w}\left(\bm{p}-\frac{1}{2}\widetilde{\bm{p}},\bm{q}+\frac{1}{2}\widetilde{\bm{q}}\right).
  \end{aligned}
\end{equation}

 The first example is for the observable $O_{\rm w}=q_1$ in an unbounded phase space.
 Direct calculation gives the following result,
\begin{equation}
  J=(2\pi\hbar)^d \left[q_1^2\delta^d(\bm{p}')\delta^d(\bm{q}')+\frac{\hbar^2}{4}\delta''(p'_1)\delta^d(\bm{q}')\prod_{\mu=2}^d\delta(p'_\mu)\right].
\end{equation}

\begin{figure}[!t]
  \centering
  \includegraphics[width=1\linewidth]{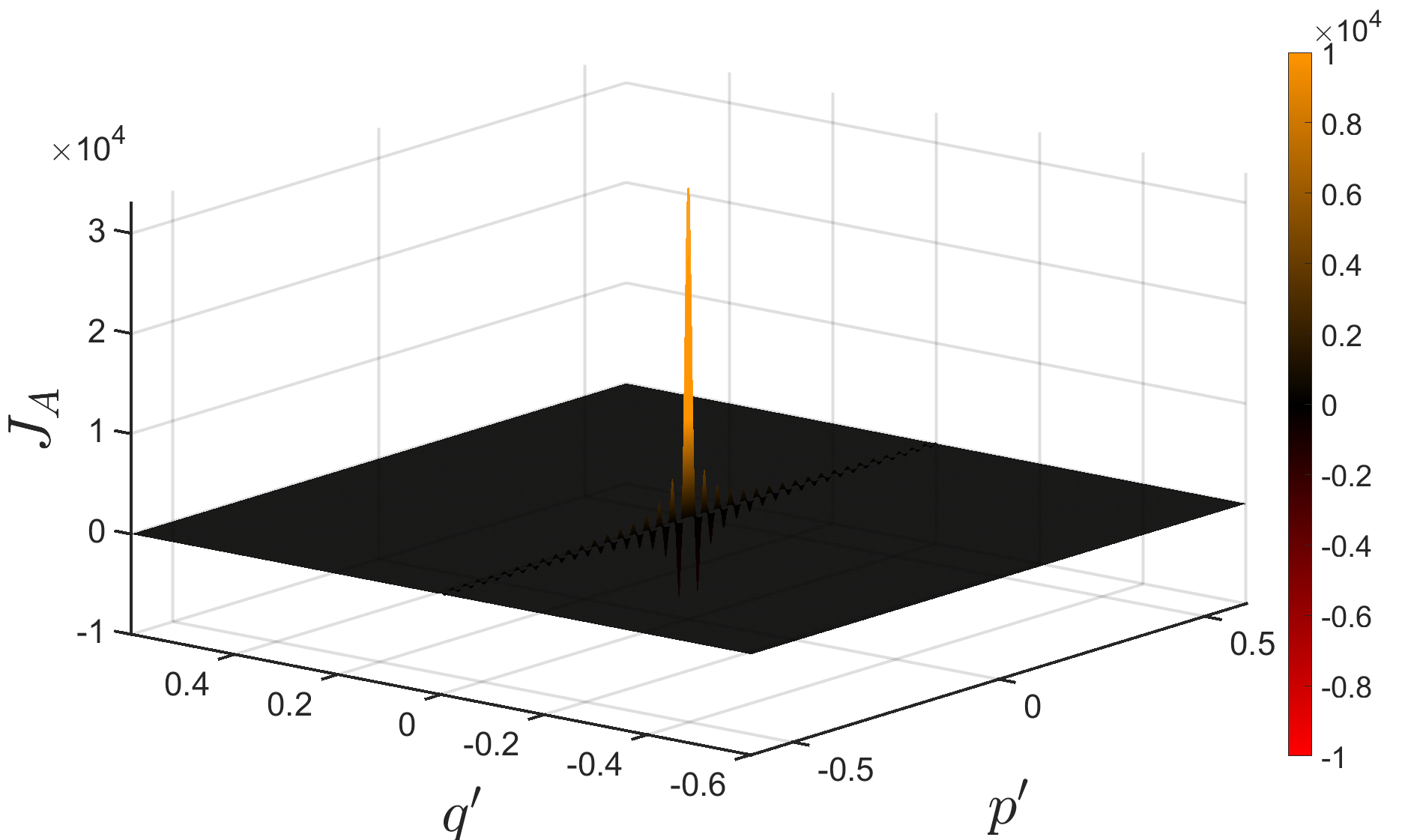}
  \caption{The image of $J_A$ given in Eq.(\ref{JA_Example}). Here the $p, q$ are fixed
  as $p=0$, $q=0$, and $\hbar=0.01$.}
  \label{JAPlot}
\end{figure}

 The second one is an observable $O_{\rm w}=p_1$, but
 due to specific dynamics of the model, the phase space of the classical counterpart of
 the model is effectively bounded. Actually, most models within specific calculations have bounded phase spaces.
 For $-1\leqslant p_\mu\leqslant 1$ and $-1\leqslant q_\mu\leqslant 1$, we found that
\begin{equation}
  J= (2\pi\hbar)^{d-1} J_A \prod_{\mu=2}^d\delta(p'_\mu)\delta(q'_\mu),
\end{equation}
where
\begin{equation}\label{JA_Example}
  \begin{aligned}
      J_A=& 2\pi\hbar\left\{\frac{1}{\pi q'_1}
      \left(\left[\left(2\abs{p_1}-1\right)+\frac{\hbar^2}{2q^{'2}_1}\right]\sin\left[2g(p_1)\frac{q'_1}{\hbar}\right]\right.\right.\\
      &\left.\left.-\frac{\hbar}{q'_1}g(p_1)\cos\left[2g(p_1)\frac{q'_1}{\hbar}\right]\right)
      \frac{1}{\pi p'_1}\sin\left[2g(q_1)\frac{p'_1}{\hbar}\right]\right\}.\\
  \end{aligned}
\end{equation}
 Here, $g(x):=1-\abs{x}$.
 The shape of $J_A$ computed from Eq.(\ref{JA_Example}) is shown in Fig.\ref{JAPlot},
 where it is seen that $J_A$ has large values only for small $\bm{p}'$ and $\bm{q}'$,
 with a pronounced peak centered at $\bm{p}'=\bm{q}'=0$.
 This point is consistent with the following relatively simple expression obtained
 in the semiclassical limit of $\hbar\rightarrow 0$, i.e.,
\begin{equation}
  \lim_{\hbar \to 0} J_A=2\pi\hbar\left[p_1^2\delta(q_1)+\frac{\hbar^2}{4}\delta''(q_1)\right]\delta(p_1),
\end{equation}
 where the following relation has been used,
\begin{equation}
  \lim_{\hbar\rightarrow 0}\frac{1}{\pi a}\sin\left[2g(b)\frac{a}{\hbar}\right]=\delta(a).
\end{equation}

As the third example, we discuss $O_{\rm w}=\bm{p}\cdot\bm{q}$, which is defined in a
bounded phase space satisfying
$-1\leqslant p_\mu\leqslant 1$ and $-1\leqslant q_\mu\leqslant 1$ for all $\mu$.
 A direct calculation gives the following corresponding $J$,
\begin{widetext}
\begin{equation}
  \begin{aligned}
      J=&(2\pi\hbar)^d \sum_{\mu=1}^d\left\{\frac{1}{\pi q'_\mu}
      \left(\left[\left(2\abs{p_\mu}-1\right)+\frac{\hbar^2}{2q^{'2}_\mu}\right]\sin\left[2g(p_\mu)\frac{q'_\mu}{\hbar}\right]
      -\frac{\hbar}{q'_\mu}g(p_\mu)\cos\left[2g(p_\mu)\frac{q'_\mu}{\hbar}\right]\right)
      \right\}\\
      &\qquad\qquad\times\frac{1}{\pi p'_\mu}
      \left(\left[\left(2\abs{q_\mu}-1\right)+\frac{\hbar^2}{2p^{'2}_\mu}\right]\sin\left[2g(q_\mu)\frac{p'_\mu}{\hbar}\right]
      -\frac{\hbar}{p'_\mu}g(q_\mu)\cos\left[2g(q_\mu)\frac{p'_\mu}{\hbar}\right]\right)\\
      &\left.\qquad\qquad\times\prod_{\nu\neq\mu} \frac{1}{\pi q'_\nu}\sin\left[2g(p_\nu)\frac{q'_\nu}{\hbar}\right]
        \frac{1}{\pi p'_\nu}\sin\left[2g(q_\nu)\frac{p'_\nu}{\hbar}\right]\right\}\\
      &+(2\pi\hbar)^d\sum_{\mu=1}^d\sum_{\nu\neq\mu}\Re\left\{F(q'_\mu,p_\mu)F^*(p'_\mu,q_\mu)F^*(q'_\nu,p_\nu)F(p'_\nu,q_\nu)\right\}
      \prod_{\lambda\neq\mu,\nu}\left\{\frac{1}{\pi q'_\lambda}\sin\left[2g(p_\lambda)\frac{q'_\lambda}{\hbar}\right]
      \frac{1}{\pi p'_\lambda}\sin\left[2g(q_\lambda)\frac{p'_\lambda}{\hbar}\right]\right\}
  \end{aligned}
\end{equation}
where
\begin{equation}
  \begin{aligned}
      F(a,b)=\frac{1}{\pi a}&\left\{g(b)\cos\left[2g(b)\frac{a}{\hbar}\right]\right.
      \left.-\left(\frac{\hbar}{2a}+ib\right)\sin\left[2g(b)\frac{a}{\hbar}\right]\right\}.
  \end{aligned}
\end{equation}
 In the semiclassical limit of $\hbar\rightarrow 0$, one finds that
\begin{equation}
  \lim_{\hbar \to 0} F(a,b)=\frac{\hbar}{2}\delta'(a)-ib\delta(a),
\end{equation}
and
  \begin{equation}
  \begin{aligned}
  \lim_{\hbar \to 0}   J&=(2\pi\hbar)^d\sum_{\mu=1}^d\left\{\left[p_\mu^2\delta(q_\mu)+\frac{\hbar^2}{4}\delta''(q_\mu)\right]\right.
    \left.\left[q_\mu^2\delta(p_\mu)+\frac{\hbar^2}{4}\delta''(p_\mu)\right]
    \prod_{\nu\neq\mu}\left[\delta(p_\nu)\delta(q_\nu)\right]\right\}\\
    &+(2\pi\hbar)^d\sum_{\mu=1}^d\sum_{\nu\neq\mu}\Re\left\{F(q'_\mu,p_\mu)F^*(p'_\mu,q_\mu)F^*(q'_\nu,p_\nu)F(p'_\nu,q_\nu)\right\}
      \prod_{\lambda\neq\mu,\nu}\left[\delta(p_\lambda)\delta(q_\lambda)\right].
  \end{aligned}
\end{equation}
\end{widetext}

\section{A $\hbar$-scaling property of $J$}\label{J_Scaling}

In this section, we give a simple scaling analysis of the bilinear phase-space convolution
function $J$,
which has been used in the main text, showing a $\rho_{\rm dos}$-scaling
behavior of the offdiagonal elements in the ETH ansatz.

For this purpose, we perform
a coordinates transformation $(\widetilde{\bm{Y}}_1,\widetilde{\bm{Y}}_2):=(\widetilde{\bm{p}},\widetilde{\bm{q}})/\hbar$
in the definition of $J$ (Eq.(\ref{J_Definition})). Within the variables $(\bm{Y}_1,\bm{Y}_2)$,
$J$ has the following form:
\begin{equation}\label{J_VariableY}
  \begin{aligned}
    J&\left(\bm{p},\bm{q},\bm{p}',\bm{q}'\right)
    =\left(\frac{\hbar}{2\pi}\right)^d\int d\widetilde{\bm{Y}}_1 d\widetilde{\bm{Y}}_2
    \ e^{-i\widetilde{\bm{Y}}_1\cdot\bm{q}'}e^{-i\widetilde{\bm{Y}}_2\cdot\bm{p}'}\\
    &\times O_{\rm w}\left(\bm{p}+\frac{\hbar}{2}\widetilde{\bm{Y}}_1,\bm{q}-\frac{\hbar}{2}\widetilde{\bm{Y}}_2\right)
    O_{\rm w}\left(\bm{p}-\frac{\hbar}{2}\widetilde{\bm{Y}}_1,\bm{q}+\frac{\hbar}{2}\widetilde{\bm{Y}}_2\right).
  \end{aligned}
\end{equation}
According to Eq.(\ref{Osc_Ocl_Error}), $O_{\rm w}=O_{\rm cl}+{\rm O}(\hbar^m)\ (m\geqslant 1)$.
Further, considering $O_{\rm cl}$ is usually a smooth function, we can expand it into a series of $\hbar$:
\begin{subequations}
  \begin{align}
    O_{\rm cl}\left(\bm{p}+\frac{\hbar}{2}\widetilde{\bm{Y}}_1,\bm{q}-\frac{\hbar}{2}\widetilde{\bm{Y}}_2\right)
    =O_{\rm cl}\left(\bm{p},\bm{q}\right)+{\rm O}(\hbar^m)\\
    O_{\rm cl}\left(\bm{p}-\frac{\hbar}{2}\widetilde{\bm{Y}}_1,\bm{q}+\frac{\hbar}{2}\widetilde{\bm{Y}}_2\right)
    =O_{\rm cl}\left(\bm{p},\bm{q}\right)+{\rm O}(\hbar^m),
  \end{align}
\end{subequations}
where $m\geqslant 1$. Substituting these into Eq.(\ref{J_VariableY}), we get
\begin{equation}
  \begin{aligned}
    J
    =&\left(\frac{\hbar}{2\pi}\right)^d\int d\widetilde{\bm{Y}}_1 d\widetilde{\bm{Y}}_2
    \ e^{-i\widetilde{\bm{Y}}_1\cdot\bm{q}'}e^{-i\widetilde{\bm{Y}}_2\cdot\bm{p}'}\\
    &\qquad\qquad\times\left[O^2_{\rm cl}\left(\bm{p},\bm{q}\right)+{\rm O}(\hbar^m)\right]\\
    =&(2\pi\hbar)^d O^2_{\rm cl}\left(\bm{p},\bm{q}\right)\delta^d(\bm{p}')\delta^d(\bm{q}')
    +{\rm O}(\hbar^{d+m}).
  \end{aligned}
\end{equation}
Obviously, due to the smallness of $\hbar$, the first item dominates $J$'s behavior.
Therefore, $J$ scales as $\hbar^d$.

\bibliography{Merge.bbl}

\end{document}